\documentclass{JHEP3}
\def\beq{\begin{equation}}
\def\endeq{\end{equation}}
\def\bea{\begin{eqnarray}}
\def\endea{\end{eqnarray}}

\def\lto{\mathop
        {\hbox{${\lower3.8pt\hbox{$<$}}\atop{\raise0.2pt\hbox{$\sim$}}$}}}

\title{Note on Bound States and the Bekenstein Bound}

\author{Donald Marolf\\
Physics Department, UCSB, Santa Barbara, CA 93106.}

\author{Radu Roiban\\
Physics Department, UCSB, Santa Barbara, CA 93106.}

\abstract{In this brief note we draw attention to examples of
quantum field theories which may hold 
interesting lessons for attempts to devise a precise formulation of
the Bekenstein bound.  Our comments mirror the recent results of 
Bousso (hep-th/03110223) indicating that the species problem remains
an issue for precise formulations of this bound.}

\date{June, 2004}


\begin{document}

\section{Introduction}

There has been much discussion in the literature of the idea that
quantum systems may be subject to certain fundamental bounds relating
their entropy ($S$) to their size (measured in terms of a radius $R$
or an enveloping area), and perhaps to their energy ($E$).  Such
proposed bounds include the Bekenstein bound $S < \alpha RE$
\cite{Bek,erice}, the holographic bound $S < A/4\ell_p^2$ 
\cite{LS,tHooft}, and the more subtle Causal \cite{BV}
and Covariant \cite{CEB} Entropy Bounds.  Such bounds 
were originally motivated by considerations of black hole
thermodynamics \cite{Bek,erice,LS,tHooft}.  Though this motivation 
has been criticized by various authors \cite{UW,MS1,MS2,MMR}, the
proposed bounds remain interesting topics of discussion and 
investigation.

The Covariant Entropy Bound
represents a refinement of the holographic bound as, at
least when spacetime can be treated classically, it gives a precise
definition of what is meant by the area $A$.  Similarly, the
parameters playing the role of size and energy for the Causal Entropy
Bound are well-defined in this context, though the same is not true of
the original holographic bound.
It is also of interest to study whether a more precise conjecture can
be found to replace the Bekenstein bound $S < \alpha RE$. 
This was explored in two recent papers \cite{RB1,RB2} by Bousso.
The Bekenstein bound is unique among those above in that it does not
involve the Planck length.  It may therefore be conjectured to 
hold in ordinary field theories, without considering coupling to
gravity.
This is advantageous for testing the bound, 
as we have more knowledge as to which such theories exist than we
do when gravity is considered. An alternate interpretation of the Bekenstein 
bound is that, although it does not explicitly refer to the Planck
length, it should apply only to field theories which can in principle 
be consistently coupled to the gravitational field.  We shall have
little to say here about this more restrictive conjecture.

In \cite{RB1} (following Bekenstein \cite{B1,B2,B3})
it was argued that a precise version of this conjecture might apply to
arbitrary quantum field theories. 
In particular, it was argued that a more precise formulation might be
able to handle the so-called `species problem', 
referring to the fact that naive interpretations of the bound $S <
\alpha RE$ (where $\alpha$ is a fixed constant of order $1$)  
are readily violated in any theory containing a large number of
fields.  A simple example arises from a one-particle wavepacket
state in a theory of $N$ massless scalar fields.  Such a state has $RE
\sim 1$, but $S \sim \ln N$.  Thus, the most naive interpretation of
the Bekenstein bound is violated. 

Bekenstein has long argued that the bound should not apply to such
wavepacket states (which will eventually spread out in space), but
only to `complete systems' \cite{B1,B2,B3} which are
truly confined to a finite region and that one should 
include contributions from the energy of any `walls' used to hold the
system together.  It is here that some cleverness is needed 
to make this statement precise since in flat spacetime, even if walls
are introduced, the full system (including the walls) will necessarily
possess an overall center of mass degree of freedom which will be
unconfined and will eventually spread out across all of space. 
Thus, it is not clear in what sense any sub-system of the universe is
truly `complete' in this sense. 

The final section of \cite{RB1} suggested that one should simply disregard the
overall center of mass degree of freedom and instead 
consider `bound states,' with the size $R$ being the width of the
bound state.  Following \cite{RB1}, we shall 
not yet be too precise about how this width is defined.  We also note
that another proposal was explored in 
\cite{RB2}, in which the conjecture was
made precise in the context 
of discrete light cone quantum field theory, where the size is
controlled by the size of a compact direction in the spacetime. 
However, it was noted in \cite{RB2} that a large number of species can violate
the bound as easily in this second context as in the naive example above. 

Here we consider the `bound state' proposal of \cite{RB1} to 
remove the center of mass 
degree of freedom and define the resulting quotient to be the set of
bound states.  One may then test bounds of the form $S < \alpha RM$,
where $M$ is the mass of the bound states.    Even in this context one
may quickly construct counter-examples.  First, consider again a theory of $N$ free massless scalar
particles. The above quotient of the one-particle Hilbert space leaves
an $N$-dimensional vector space corresponding to particle
flavor.  But $S = \ln N$ and $M=0$ clearly violates the bound, and at
finite $M$ the bound is violated for large enough $N$. This
counter-example was also pointed out in \cite{RB2}.     A similar
trivial violation may be constructed from any theory having a clear
`bound state' (say, QCD with its hadrons) and then considering a
Lagrangian built from a large number $N$ of mutually non-interacting
copies of this system.   

Now, while the examples above are counter-examples in the technical
sense, they appear to be somewhat trivial.  This triviality might be
taken as an indication that, with a bit of refinement, the bound state
version of the Bekenstein bound could be made more robust. 
Our purpose below is to point out less trivial counter-examples in
which the states appear at some intuitive level to all be `bound states
held together by the same force' -- though the existence of dual
formulations again raises the question of to what extent `bound
states' are fundamentally different from any other sort of state. 
Our counter-examples concern 
${\cal N}=1$ and ${\cal N}=2$ supersymmetric $SU(N_c)$ gauge theory,
where the degrees of freedom are under some control (see for example
\cite{INSE}) in the infrared limit.   

\section{Examples: ${\cal N}=1$ and ${\cal N}=2$ $SU(N_c)$ gauge theory with 
fundamental matter}

Let us consider an $SU(N_c)$ gauge theory in 3+1 dimensions with matter in the
fundamental representation. The infrared behavior depends on the
number $N_f$ of matter 
multiplets $Q$ and ${\tilde Q}$ (see for example \cite{INSE}). Quite
generally, if $N_f< 3N_c$ among the low energy degrees
of  freedom one finds, in some description, $N_f^2$ mesons 
${\cal M}=Q{\tilde Q}$ and 
baryons, which are composites of the high energy matter fields. 
Of particular interest to our discussion is the situation
$\frac{3}{2}N_c<N_f<3N_c$ in which the theory has a nontrivial 
infrared fixed point; i.e., the theory flows to a conformal field
theory in the infrared\footnote{The cases with no conformal invariance
are quite hard to analyze because the masses of the low energy degrees of
freedom depend on the K\"ahler potential and are not under control.}. 
Since ${\cal M}$ are chiral fields, the superconformal algebra relates their
dimension to their ${\cal R}$ charge through 
\begin{equation}
{\cal D}({\cal M})=\frac{3}{2}{\cal R}({\cal M})=3\frac{N_f-N_c}{N_f}~~,
\end{equation} 
which is less than the sum of the dimensions of the constituents, and
thus one can think of the mesons as bound states. Since the  
theory is conformal, it is clear that the mesons are exactly massless and that
at long wavelengths they may be described as having zero
size\footnote{The other conformally invariant answer would be infinite
size, but conformal invariance is broken
away from the fixed point so that the mesons will have some finite
size in the full theory (which can then be neglected in the long
wavelength limit).}.  We 
note that for {\it any} $N_f, N_c$ in our allowed range, counting the
meson degrees of freedom as bound states with $E=M =0$ violates $S <
\alpha RE$ in the sense described above, 
even though the number of mesons is sometimes lower than the number of
fundamental fields (for $\frac{3}{2}N_c < N_f < 2 N_c$).  Furthermore,
even if one changes the rules and requires that the mesons be placed
in wavepackets with $E \sim 1/R$, the bound is readily violated at
large $N_c,N_f$. 

In the limit $N_f\searrow \frac{3}{2}N_c$ the dimension of the
meson fields becomes unity which leads to them being interpreted as
free fields. We thus find the `species problem' in a naively
interacting field theory. One might be tempted to discard this
limiting case based on the existence of a dual formulation in which
the theory is free; it is however not clear why one should do this at
intermediate energies.

Another class of  examples in the direction described above is
provided by ${\cal 
N}=2$ theories with $SU(N_c)$ gauge group and $N_f=2N_c$ fundamental
matter fields. These theories are also conformal and the discussion is
quite similar to the case of ${\cal N}=1$ theories, so we will not repeat it.

\section{Brief Discussion}

\label{disc}

We have drawn attention to a set of models which contain a large number of
massless states associated with composite 
operators of more fundamental fields.  In this sense, these states are
bound and would violate a `bound state version' of 
a Bekenstein bound.  In fact, even a single massless bound state
violates a strict interpretation of this bound.  We therefore note
that proponents of a 
bound-state version of a Bekenstein-like bound must advocate 
either 1) application of such bounds only to theories which can be
consistently coupled to gravity or 2) a notion of bound state 
for which the above theories fail to qualify.   Note that option (1)
would be ruled out if a convincing theory of the above models 
coupled to gravity could be found.
We mention in passing that 
similar results hold for $SU(n)$ gauge theory with $m< n$ massless
fundamental fermions and ${\cal N}=2$ supersymmetry in 3+1 dimensions which
are confining rather than conformal.  Here the mass of mesons can be
tuned to be arbitrarily small while taking the confinement energy
scale to infinity as fast as one likes, and thus presumably keeping
the size of the meson bound states small.

Let us return briefly to the actual context discussed in \cite{RB1},
in which Bousso attempted to study the confinement of 
degrees of freedom to a fixed region of space through the use of an
external potential.  The argument in \cite{RB2} 
was that an attempt to use a single potential to confine a large
number of species inevitably leads to large radiative 
corrections, over which one has little control.  The bound states in
the models above are of a similar nature, as the gluons couple equally
to each of the $N_f$ flavors of $Q$ and $\tilde Q$, though now
supersymmetry  
{\it does} allow one to retain some control over the analysis.  In the
context of such bound states one finds that 
degrees of freedom (the relative motion of the constituents) can in
fact be confined without great cost in energy. 
This suggests that if tools could be found to make the analysis
tractable, external potential problems of the sort 
studied in \cite{RB1} could also lead to large numbers of states
localized in a region of fixed size.   As usual, we expect that
strongly coupled 
quantum field theories are capable of all manner of 
surprising behaviors not immediately obvious from their perturbative
description. 

\medskip

{\bf Acknowledgments:} 
We are grateful to Tom Banks, Jan de Boer, and especially Raphael Bousso for several interesting discussions
on these issues. D.M. was supported in part by NSF grant PHY03-54978,
and by funds from the University of California and the Kavli Institute
of Theoretical Physics.  R.R. was supported in part by the National
Science Foundation under Grant No.~PHY00-98395 as well as by the
Department of Energy under Grant No.~DE-FG02-91ER40618.

\end{document}